\documentclass[11pt]{article}
\usepackage{times}
\input amssym.def
\input amssym.tex
\newcommand{\hsp}{\hspace*{\parindent}}

\catcode`\@=11
\renewcommand{\section}{
	\setcounter{equation}{0}
	\@startsection {section}{1}{\z@}{-3.5ex plus -1ex minus
	-.2ex}{2.3ex plus .2ex}{\large\bf}
	}
\catcode`@=12

\makeatletter
\def\@sect#1#2#3#4#5#6[#7]#8{\ifnum #2>\c@secnumdepth
     \def\@svsec{}\else 
     \refstepcounter{#1}\edef\@svsec{\csname the#1\endcsname.\hskip .75em }\fi
     \@tempskipa #5\relax
      \ifdim \@tempskipa>\z@ 
        \begingroup #6\relax
          \@hangfrom{\hskip #3\relax\@svsec}{\interlinepenalty \@M #8\par}%
        \endgroup
       \csname #1mark\endcsname{#7}\addcontentsline
         {toc}{#1}{\ifnum #2>\c@secnumdepth \else
                      \protect\numberline{\csname the#1\endcsname}\fi
                    #7}\else
        \def\@svsechd{#6\hskip #3\@svsec #8\csname #1mark\endcsname
                      {#7}\addcontentsline
                           {toc}{#1}{\ifnum #2>\c@secnumdepth \else
                             \protect\numberline{\csname the#1\endcsname}\fi
                       #7}}\fi
     \@xsect{#5}}
\def\@begintheorem#1#2{\it \trivlist \item[\hskip \labelsep{\bf #1\ #2.}]}
\makeatother
\makeatletter
\def\eqalignno#1{\displ@y \tabskip\@centering
  \halign to\displaywidth{\hfil$\@lign\displaystyle{##}$\tabskip\z@skip
    &$\@lign\displaystyle{{}##}$\hfil\tabskip\@centering
    &\llap{$\@lign##$}\tabskip\z@skip\crcr
    #1\crcr}}
\makeatother

\begin{document}
\begin{center}
{\Large\bf Competition and cooperation:  Libraries and publishers in the transition to electronic scholarly journals} \\
\vspace{\baselineskip}
Andrew Odlyzko \\
\vspace{0.5\baselineskip}
AT\&T Labs - Research \\
Florham Park, New Jersey 07932 \\
amo@research.att.com \\
\vspace{\baselineskip}
January 19, 1999 \\
\vspace{1.5\baselineskip}
{\bf ABSTRACT}
\vspace{.5\baselineskip}
\end{center}
\setlength{\baselineskip}{1.5\baselineskip}

The conversion of scholarly journals to digital format is proceeding
rapidly, especially for those from large commercial and learned
society publishers.  This conversion offers the best hope for survival
for such publishers.  The infamous ``journal crisis'' is more of a
library cost crisis than a publisher pricing problem, with internal
library costs much higher than the amount spent on purchasing books
and journals.  Therefore publishers may be able to retain or even
increase their revenues and profits, while at the same time providing
a superior service.  To do this, they will have to take over many of
the function of libraries, and they can do that only in the digital
domain.  This paper examines publishers' strategies, how they are
likely to evolve, and how they will affect libraries.

\clearpage
\thispagestyle{empty}
\setcounter{page}{1}
\section{Introduction}
\hsp
A specter is haunting the publishing industry.  It is the specter of
{\em Encyclopaedia Britannica.}  My first paper on electronic publishing
\cite{Odlyzko1} cited {\em Encyclopaedia Britannica} as an example of a
formerly flourishing business that fell into trouble in just a few
years by neglecting electronic media.  Since that time, {\em Encyclopaedia
Britannica} has collapsed, and was sold to Jacob Safra, who is
investing additional funds to cover losses and revamp the business
\cite{Melcher}.  The expensive sales force has been dismissed, and while
print versions can still be purchased from bookstores, the focus is on
electronic products.  This collapse occurred even though
{\em Encyclopaedia Britannica} had more than two centuries of tradition
behind it, and was by far the most scholarly and best known of the
English-language encyclopedias.  In the apt words of \cite{EvansW},
\begin{quote}
  Britannica's downfall is more than a parable about the dangers of
  complacency.  It demonstrates how quickly and drastically the new
  economics of information can change the rules of competition,
  allowing new players and substitute products to render obsolete
  such traditional sources of competitive advantage as a sales force,
  a supreme brand, and even the world's best content.
\end{quote}
This paper concentrates on scholarly journals.  Not only that, but it
will not deal with journals such as {\em Science} or {\em IEEE Spectrum,} which
are distributed to tens or hundreds of thousands of readers.  It will
concentrate on the low-circulation journals that are sold primarily to
libraries, and typically have about a thousand subscribers.  These are
the journals that bring in the bulk of revenues to scholarly
publishers, and are the source of the research library crisis.  Still,
the {\em Encyclopaedia Britannica} example will be used several times in
analyzing these journals.  The markets are different, but there are
many similarities.

A few years ago there was considerable skepticism whether electronic
journals were feasible at all.  A large part of \cite{Odlyzko1} was
therefore devoted to demonstrating that Licklider \cite{Licklider} was
right in the early 1960s in predicting that by the late 1990s,
computing, communications, and storage technologies would be adequate
for handling the scholarly literature.  By now, most such doubts have
been dispelled (although there are still exaggerated concerns about
durability of digital storage as well as technical standards).  It is
also widely accepted that electronic journals are desirable and
inevitable.  Therefore we see rapid growth of digital material.
Scholarly journals that exist only in electronic formats continue to
proliferate.  However, since they started from a low base, they still
cover a small fraction of the literature.  The dominant electronic
journals (if not in absolute numbers, then certainly in amount of
peer-reviewed material) are digital versions of established print
serials.  (See \cite{ARL,HitchcocCH} for latest estimates of the
electronic marketplace.)  The largest scholarly publisher, Elsevier,
will soon have all its approximately 1200 journals available
electronically.  Professional societies, such as the ACS, APS, AMS,
and SIAM, also have either already created electronic versions of all
their research journals, or are in the process of doing so.  The
question of whether most scholarly journals will be electronic or not
is thus settled.

While it is now widely accepted that scholarly journals have to be
electronic, how they are to be delivered, and especially at what
price, remains to be decided.  This article examines the current
practices by publishers, both commercial and professional society
ones, and their likely evolution and impact on libraries.

Some features of the electronic offerings from established publishers
(such as offering only bundles of journals, without a chance to
purchase individual ones) are causing controversy among scholars and
librarians.  The subtitle of the article \cite{Kiernan1} describes the
mixture of reactions well:  ``Some see a way to meet professors' needs;
others say publishers are protecting profits.''  There is no doubt that
the publishers' primary motive is protection of revenues and profits.
This is true for both commercial and learned society publishers.
Still, this article argues that professors' needs are likely to be
better satisfied by these new electronic offerings than by traditional
print journals.  However, for the publishers to protect their revenues
and profits, they will have to usurp much of the role and resources of
libraries.  Further, publishers' success is likely to retard the
development of an even more efficient system.

{\em Encyclopaedia Britannica} was vulnerable largely because it had an
enormously bloated cost structure.  The \$1,500 to \$2,500 that
purchasers paid for each set included a couple of hundred dollars for
the printing, binding, and distribution.  Most of the rest was for the
sales force and general administrative overhead.  The vaunted
editorial content apparently amounted to well under 10 percent of the
total price.  That is what allowed \$50 CD-ROM encyclopedias to
compete.  They did not have the same quality of content, nor the
nicely printed volumes, but they did have superior searchability,
portability, and an irresistible price.

It is important to note that after some abortive attempts to sell
first \$1,200, then \$300 CD-ROMs, {\em Encyclopaedia Britannica} is now
offering its CD-ROMs for \$125 or even less.  It is not known publicly
what its total budget or internal cost allocations are, but it appears
safe to say that the entire encyclopedia industry is spending much
more on content than it used to.  At Britannica, editorial staff
reportedly has increased by over 25 percent.  Further, usage of
encyclopedias has probably increased substantially.  While most of the
CD-ROM versions are hardly ever used (which was also true of the paper
editions, of course), there are tens of millions of them, many more
than the print encyclopedias.  This means that total usage is surely
up.  Universities that subscribe to the online version of the
{\em Encyclopaedia Britannica} report that usage is far higher than it
ever was for the printed versions \cite{Getz}.

As with {\em Encyclopaedia Britannica}, the main effect of new
technologies on other parts of the publishing industry will be
elimination of costs that once were unavoidable.  Spending on content
will go up.  Total profits, which many finger as the culprit in the
library crisis, may also increase.  (It was noted in \cite{Odlyzko1} that
while revenues of the {\em World Book} encyclopedia went down when it
switched to a CD-ROM format, profits grew.)  However, the entire
information industry is likely to become much more efficient, with
more resources devoted to the intellectual content that matters.

The current scholarly journal system is full of unnecessary costs.
The ones that have attracted the most attention in the past were those
associated with publishing.  The main traditional functions of
publishers, in which they handled copy editing, production, and
distribution of material provided to them for free by scholars, are
mostly obsolete.  The difference in quality between the manuscripts
that scholars can produce themselves, and the final printed journal
versions, has decreased almost to the vanishing point with the arrival
of easy to use computerized typesetting.  (Here I am referring to copy
editing and other tasks performed by professionals at publishers.
Peer review is another matter.  It was and continues to be done gratis
by scholars, so that even if it is facilitated by publishers today, it
can be performed without them.)

To a large extent publishers are responding to cuts in subscriptions
of large (and therefore expensive) journals by launching smaller, more
specialized serials.  These are often treated with much less care, so
they are not much better in quality of presentation than camera-ready
journals.  Furthermore, they often have laughably small circulations
(such as the figure of 300 or lower cited by a publisher \cite{Beschler}).
Thus the current scholarly journal system is becoming dysfunctional.

To survive in the long run, publishers will need to move towards
provision of intellectual value (such as that provided by the staffs
of reviewing journals).  That is a hard task, requiring new skill
sets, and often new personnel.  What keeps the publishers' situation
from being hopeless is the tremendous inertia of the scholarly
community, which impedes the transition to free or inexpensive
electronic journals.  Another factor in the publishers' favor is that
there are other unnecessary costs that can be squeezed, namely those
of the libraries.  Moreover, the unnecessary library costs are far
greater than those of publishers, which creates an opportunity for the
latter to exploit and thereby to retain their position.

Section 2 briefly reviews the economics of scholarly journals.
Section 3 discusses the basic strategy that established publishers are
following in moving to electronic journals.  Section 4 concentrates on
some features of the current electronic journal pricing and licensing
policies.  Finally, Section 5 offers some speculation about the
future.

\section{Economics and technology}
\hsp
This section reviews the basic economic facts about scholarly journal
publishing.  They were first presented in \cite{Odlyzko1} and then in
greater detail (and with more data about electronic journals, based on
more experience) in \cite{Odlyzko4}.  See also \cite{TenopirK}.

Conventional print journals bring in total revenues to publishers of
about \$4,000 per article.  On the other hand, there are many
flourishing electronic journals that operate without any money
changing hands, through the unpaid labor of their editors (and with a
trivial implicit subsidy by the editors' institutions that provide
computers and network connections).  There is still some question
whether this model can scale to cover most of peer-reviewed literature
and satisfy scholar's needs.  Even if the totally free journals will
not suffice, experience has shown that quality that is perfectly
adequate for most readers can be produced in the electronic
environment for less than \$400 per article \cite{Odlyzko4}.  Such costs can
be recovered either through subscription fees or charges to authors,
and both models are being tried.

Journal subscription costs are only one part of the scholarly
information system.  As was pointed out in \cite{Odlyzko1}, internal
operating costs of research libraries are at least twice as high as
their acquisition budgets.  Thus for every article that brings in
\$4,000 in revenues to publishers, libraries in aggregate spend \$8,000
on ordering, cataloging, shelving, and checking out material, as well
as on reference help.  The scholarly journal crisis is really a
library cost crisis.  If publishers suddenly started to give away
their print material for free, the growth of the literature would in a
few years bring us back to a crisis situation.

It is important to emphasize the point about the cost of libraries.
The \$4,000 per article is rough estimate (see \cite{Odlyzko1,Odlyzko4,TenopirK}) and one can argue that the precise figure should be higher
or lower.  On the other hand, the exact dollar figures for the 120
members of the Association of Research Libraries, which includes most
of the large research libraries in the U.S. and Canada, do show that
purchases of books, journals, and other materials make up rather
consistently about a third of their budgets, and have done so for
years \cite{ARL}.  The other two thirds goes overwhelmingly for salaries
and wages of librarians and support staff, with a small fraction for
items such as binding.  The table below shows the breakdown of library
expenditures at several universities during the 1996--97 academic year,
taken from the comprehensive statistics collected by the ARL and
available online at \cite{ARL}.  (Harvard has the world's highest library
budget.)
\begin{center}
\begin{tabular}{lrrrr}
~ & \multicolumn{1}{c}{circulation} & \multicolumn{1}{c}{staff} & \multicolumn{1}{c}{purchases} & \multicolumn{1}{c}{total budget} \\ [+.1in]
Brown & 0.3M~~~~~ & 240~~ & \$5.0M~~ & \$14.8M~~~ \\
Harvard & 1.4M~~~~~ & 1182~~ & \$17.5M~~ & \$70.9M~~~ \\
Ohio State & 1.5M~~~~~ & 423~~ & \$8.6M~~ & \$22.1M~~~ \\
Princeton & 0.6M~~~~~ & 384~~ & \$9.2M~~ & \$24.9M~~~
\end{tabular}
\end{center}

This division of costs has held for a long time.  For example, in the
1996--97 academic year, Harvard spent 24.7\% of its library budget on
acquisitions, whereas in 1981--82 it spent 27.5\% (\$5.8M out of \$21.1M)

The ARL numbers substantially underestimate the internal costs of
libraries, since they include neither the costs of the buildings, nor
of building maintenance, nor of employee fringe benefits.  In many
cases those numbers also fail to include the costs of library
automation systems.  If those additional costs were to be included,
costs of acquisitions might turn out to be under a quarter of the
total costs of the library system \cite{Getz}.  Thus, even though much of
the cost to a library that is associated with a journal is incurred in
the future, in preserving the issues and making them accessible, it
seems safe to say that the internal costs of the library associated
with that journal are at least twice the purchase price.

The high internal costs of libraries come from the need to provide
information about, and easy access to huge collections of material
that are used infrequently at any single place.  As an example,
suppose that we ignore all the other activities of the Harvard
libraries, and allocate the entire library cost to circulating items.
We would then discover that circulating the 1.4 million items that
were borrowed (out of 13.6 million volumes in the Harvard collection
\cite{ARL}) cost around \$50 each.  By comparison, there are commercial
services (aimed at allowing publishers to reprint books in extremely
small runs) that will digitize a book for a one-time fee of \$100 to
\$150, and then print individual copies of a 300-page book for about \$5
\cite{NYT}.  That is an order of magnitude reduction in cost.  Of course,
this comparison ignores all the other function of the library, but it
does demonstrate the dramatic cost savings that are becoming possible
if one can cut back on the acquisition and management of a physical
collection.

The high cost of operating libraries is giving publishers a chance to
maintain their revenues.  Standing at the level of \$4,000 articles,
they are naturally reluctant to jump into the chasm of free or at most
\$400 articles.  Instead, they are enviously eyeing the \$8,000 per
article spent by libraries.  They are responding, either by careful
design, or through competitive instinct, in ways that should reduce
the costs of the total system by decreasing the role and cost of
libraries.  To the extent they succeed, this should produce a much
superior scholarly information system, although still an unnecessarily
expensive one.

There have been occasional proposals that libraries take over the
functions of publishers.  Given the unnecessarily high price structure
of publishers, such a course is conceivable.  However, what is much
more likely to happen in the competition for resources between
libraries and publishers is that it will be the publishers who will
come out ahead.  There are cultural, economic, technological, and
legal reasons for this prediction:
\begin{enumerate}
\item
There are fewer publishers, so it is easier for them to
mount electronic publishing efforts on a large scale,
\item
Publishers are more used to competition than librarians,
who stress cooperation,
\item
Publishers control copyrights, and thus conversion of old
material (crucial for reducing library costs) cannot be
carried out without their cooperation,

and, perhaps most important,

\item
The publishers' target is more inviting:  there is more than
twice as much resources for them to go after as there is
for librarians.
\end{enumerate}

If the scholarly publishing business were efficient and run for the
benefit of the scholarly enterprise, both libraries and publishers
would have to shrink rapidly.  However, this business is anything but
efficient.  A major contributor to this inefficiency is academic
inertia.  As shown in the discussion of rates of change in \cite{Odlyzko6},
academia is among the slowest to change in general.  Further,
scholarly publication is a sufficiently small part of research life
that it does not attract much attention.  Libraries usually consume
3\% to 4\% of university budgets, so any savings that might be realized
from library cutbacks would not make a dramatic difference to total
spending.  (Among the academic ARL members, library spending averages
about \$12,000 per full time faculty member \cite{ARL}.)  Furthermore,
library buildings, often the most prominent on campus, easily attract
donors who like to see their names immortalized on such central
facilities.

The most convincing demonstration of scholarly inertia is the reaction
(or the lack of it) to the Ginsparg preprint archive.  Starting in
1991, it has become the fundamental communication method for a growing
roster of fields, starting with theoretical high energy physics, later
spreading to other areas of physics, and now also to computer science
and mathematics \cite{Ginsparg}.  It is a sterling example of how
technology can lead to a sudden, profound, and beneficial
transformation.  Yet in 1998, this archive still processed only 24,000
submissions, which is substantial (about half of the volume of all
mathematics papers published that year), but small compared to the
perhaps 2 million papers in all STM (science, technology, medicine)
areas.  The attractions of the archive are great.  It transforms the
mode of operation of any community of scholars that embraces it, and
the transition is invariably one-way, as not a single group has
abandoned it.  It quickly becomes the dominant mode of communication
inside any group that embraces it.  However, in spite of extensive
publicity, it has not swept scholarly communication yet.  It appears
that there were special cultural factors that led to the quick
adoption of the archive by Ginsparg's own community of theoretical
high energy physicists (primarily the reliance on massive mailings of
preprints), and it has been a struggle for pioneers in other areas to
duplicate the process.  There are still many areas (especially in
chemistry and medicine) where not just preprint archives, but
preprints themselves, are rare, and in which prestigious journals get
away with policies that forbid any formal consideration of a paper
that has been circulated in preprint form.

The significance of the Ginsparg archive is two-fold.  On one hand, it
shows that scholars can embrace new technology in a short period and
derive enough benefit that giving it up becomes unthinkable.  On the
other hand, it also shows that it requires a substantial critical mass
or an external push in an area to make the transition.  In most of the
STM fields, this critical mass is not present yet.

A Ginsparg-style centralized preprint archive (or a decentralized
system like MPRESS from the European Mathematical Society) is not
compatible in the long run with an expensive journal publishing
operation that collects \$4,000 per article.  ``Available information
determines patterns of use'' in the apt words of Susan Rosenblatt
\cite{Odlyzko5}, and if the basic preprints are available for free, few
will pay a fortune for slight enhancements, which is all that current
journals offer.  The question is what is meant by ``the long run.''  The
discussion in \cite{Odlyzko6} as well as that above about the Ginsparg
archive shows that academia moves at a glacial pace.  Even in
Ginsparg's own theoretical high energy physics community, most
researchers still publish their papers in conventional print journals
(although a few senior ones have given up the practice on the grounds
that it does not serve to propagate their results).  Thus if academia
were left to itself, the current journal system might continue to
stumble along for a couple of decades until the subversive effect of
preprints would make it clear the system was not worth its cost.

In the discussion on diffusion of new technologies in \cite{Odlyzko6}, many
rapid transitions were identified with the presence of forcing agents,
people or institutions that can compel action.  The prediction of
\cite{Odlyzko1} was that a collapse of the existing print journal system
would come when academic decision makers (presidents, deans, ...)
realized that this system was superfluous, and go to departments with
offers of the type ``Would you rather stay with the existing library
system at \$12,000 per head, or would you be willing to cut that back
to \$6,000 per head, and use the savings for salaries, travel, ...?''  
I think this is still the most likely scenario for change, but that it
will involve abandonment of print and cutbacks in libraries, and less
of a cutback at publishers.  Publishers, who have been scared of
electronic publishing, are likely to become forcing agents, and speed
the transformation.

\section{The demise of print journals}
\hsp
Most established publishers have already created or are creating
electronic versions of their scholarly print journals.  Often they are
offering these digital editions at no extra cost to subscribers to the
print versions.  In some cases, institutions that forego the print
version receive a modest discount.

A coherent strategy for the publishers should contain two additional
steps in the future.  The first step is to eliminate print editions
entirely.  (This has not yet been announced by any major publisher.)  The
second one is to convert the old issues to digital form, either
themselves or through organizations like JSTOR \cite{Guthrie}.  (This is
being done by several professional society publishers, but not yet by
any commercial ones.)  This would get libraries out of the journal
distribution and archiving business (except as licensing agents, to be
discussed below) and allow for drastic reductions in library budgets.

Eliminating print editions would allow for some reduction in costs of
publishers (even if they kept their current expensive editing system),
so they have a financial incentive to do it.  In digitization, they
would have to spend money beyond their current budgets.  The key point
is that it would not be a lot of money.  An earlier article \cite{Odlyzko4}
mentioned a range of digitization costs between \$0.20 and \$2.00 per
page.  There are now projects (such as the commercial one for book
reprinting mentioned above \cite{NYT}, and the Florida Entomological
Society's project described in \cite{Walker}) that show one can obtain a
high quality digital version for \$0.60 per page.  To put these numbers
in perspective, all publishers collectively get about \$200 million per
year for mathematical journals.  On the other hand, the entire
mathematical literature accumulated over the centuries is perhaps 30 
million pages, so digitizing it
at a cost of \$0.60 per page would cost \$18 million, less than 10\% of
the annual journal bill.  Further, this would be a one-time expense.

On the way towards eliminating print editions, publishers will have to
solve a few thorny problems.  One of them is interlibrary loans.
Except for a few small organizations, until recently all publishers
had blanket prohibitions on the use of electronic editions for
interlibrary loans.  This was naturally resented by librarians, who
rely on such loans to satisfy a small but important and growing
fraction of their clients' demands.  Without the right to use
electronic editions for interlibrary loans, libraries were almost
uniformly unwilling to even consider abandoning print editions.
Recently some large publishers have announced changes in their
policies.  Electronic editions of journals of those publishers can now
be used to satisfy interlibrary loan requests, but only by printing
out the requested articles and sending them out in the printed form.
Libraries will thus have the same functionality as before (or even
better, since there will be no need to find volumes on shelves and
make photocopies).  The continued prohibition on electronic delivery
of the electronic version should suffice to maintain the distinction
between owning and borrowing that does not naturally exist in
cyberspace, and thus maintain demand for subscriptions.

Can print journals be eliminated?  Previous predictions of the eclipse
of printed matter by microfilm, for example, failed to come true.
(See \cite{Odlyzko1} for a brief survey and references to numerous faulty
predictions in this area.)  Print is certainly persistent, as has been
observed many times (cf.  \cite{Crawford}).  There is even a commercial
publisher that is about to start selling a print edition of the
{\em Electronic Journal of Combinatorics,} the most successful of the free
electronic journals in mathematics.  (The electronic version will
remain free, and the publisher will only get rights to distribute a
print version.)  Yet I am convinced that printed journals are largely
on their way out.  I do not mean that print is on its way out.  For
reasons of technology and inertia, print is likely to be with us for
several decades, and even proliferate, as personal computer printers
improve in quality and drop in price.  All that will happen is that
there will be a simple substitution, the kind that eases all
technological transitions \cite{Odlyzko6}.  Scholars will print articles on
their personal or departmental printers instead of going to the
library, photocopying those articles, and bringing the copies back to
their offices to study.

The transition to electronic distribution and storage should not take
too long.  There is tremendous inertia in academia, with some scholars
swearing that nothing can substitute for browsing of bound printed
journals.  However, this resistance can be overcome.  We already have
examples of academic libraries in which efficient document delivery
(from the library's own collections) has drastically reduced physical
visits to the library by faculty and students.  Further, network
effects will be playing an increasing role.  More material available
in electronic formats and increasing linking of digital forms of
articles will all be making it much more attractive to browse on a
screen and print out articles for careful study.  For example, in
mathematics, the two main reviewing publications, {\em Mathematical Reviews}
and {\em Zentralblatt f\"ur Mathematik,} whose electronic forms are catching
on much faster \cite{AndersonDR} (for obvious reasons of much greater
efficiency) than online versions of primary research journals, are
beginning to offer links to articles being reviewed.  Publishers will
surely help this move by making the electronic versions more
attractive than print ones.  They are already beginning to provide
links to references, and making online versions of articles available
earlier than the print editions.  At some point they will surely also
increase the prices of print editions (compared to the online ones),
and perhaps lengthen print publication backlogs.  Eventually, enough
libraries will agree to eliminate print subscriptions that they will
be phased out.  (As an intermediate step, they might be farmed out to
specialized inexpensive publishers to produce out of the electronic
versions.)  What I am predicting is that publishers, who used to
resist electronic publishing, will, out of self-interest, play the
role of the forcing agents that accelerate natural technological
transitions \cite{Odlyzko6}.

The elimination of print editions of journals will eventually reduce
publishers' costs.  (Even though they have yet to concede that
acceptable quality can be obtained in electronic publishing for 10\% of
the current print costs, they do admit that savings of 20--30\% can be
obtained by elimination of printing and distribution costs.)  Most
important, this step will reduce library costs and relieve the cost
pressures on academic information systems.  Thus the decisive steps
towards eliminating print versions of journals are likely to be taken
by academic decision makers, the deans and presidents, when they
realize how much can be saved.

What about librarians?  I expect they would adjust easily to a
paperless journal environment.  First of all, transition would be
gradual.  While there is inertia among scholars, there is also a much
more understandable inertia in the library system, given the huge
accumulated print collections.  These collections will have to be
maintained until the slow conversion to digital format is completed.
(And some materials will never be converted.)  Further, there may well
be a revival of scholarly monograph publishing, which has been getting
squeezed out of library budgets by journals.  (It is hard to forecast
what effect this will have on the libraries, though, since the number
of monographs published is likely to increase, but many of them will
be distributed electronically.)  The main job losses will be in the
less-skilled positions (with the part-time student assistants who
check out and reshelve material going first).  Reference librarians
are likely to thrive, although their job titles may not mention the
library.  After all, we will be in the Information Age, and there will
be much more information to collect, classify, and navigate.
Information specialists are likely to abound, and to have much more
interesting jobs.

Although there will be many opportunities, librarians will have to
compete to retain their preeminence as information specialists
\cite{Odlyzko5}, and operate in new ways.  However, there are two other
jobs that they are also well-positioned to retain.  One if that of
negotiating electronic access licenses.  The other is that of
enforcing access restrictions.

It is worth emphasizing that if the publishers do succeed in their
approach, and disintermediate the librarians while retaining their
revenues and profits, the resulting system is likely to be much superior 
to the present one.  Defenders of the current libraries tend to come 
from top research universities, which do have excellent library 
collections.  That is an exception, though.  Most scholars, and an 
overwhelming majority of the population, make do with very limited access 
to those precious storehouses of knowledge.  (There is an illuminating 
graph in \cite{GriffithsK}, reproduced as Fig.~9.4 on p.~202 
in \cite{Lesk}, that 
shows library usage decreasing rapidly as the effort to reach the library
grows, even on a single campus.  For the bulk of the world's
population, little is available.)  Electronic publishing promises far
wider and superior access.  I am not forecasting a new age of
universal enlightenment, with the couch potatoes starting to read
scholarly articles.  However, there will be growth in usage of
scholarly publications by the general public.  The informal
associations devoted to discussions of medical problems (those on AIDS
present the best example) show how primary research material does get
used by the wide public if it is easily available.  For scholars
alone, there will be a huge increase in productivity with much easier
access to a wider range of information.

The basic strategy of the publishers, faced with pressure to reduce
costs, is to disintermediate the libraries.  There is nothing
nefarious in this approach.  As we move towards the information age,
different groups will be vying to fill various rapidly evolving
ecological niches.  After all, many scholars are proposing that they
and the librarians disintermediate the publishers, while others would
bypass librarians and publishers both, and handle all of primary
research publishing themselves.  In this environment, some of the
potentially extremely important players might be Kinko's copy shops.
They may end up disintermediating the bookstores and libraries, by
teaming up with publishers to print books on demand.  They might also
disintermediate the publishers, by making deals directly with authors
and their agents.

\section{Fairness and the new economics of information goods}
\hsp
The previous section outlined the strategy that established publishers
appear to be pursuing or likely to pursue.  Here we discuss the
tactics.  There are extensive fears and complaints about the pricing
and access policies publishers offer for their electronic journals, as
can be seen in the messages in \cite{LIBL,NSPI}.  Many of these concerns
are likely to be allayed with time, as they are natural outcomes of a
move towards a new technological and economic environment.  By
negotiations, compromise, and experiment, librarians and publishers
will work out standard licensing terms that they and scholars can live
with.  As one example, there is great concern among librarians and
scholars about access to electronic journal articles once a
subscription is canceled.  This is clearly an issue, but one that can
be solved through negotiations.

Some issues that are raised by librarians will not go away.  The basic
problem with information goods is that marginal costs are negligible.
Therefore pricing according to costs is not viable, and it is
necessary to price according to value.  What this means is that we
will be forced into new economic models.  Many people, especially Hal
Varian \cite{Varian}, have been arguing for a long time that we will see
much greater use of methods such as bundling, differential quality,
and differential pricing.  (See also \cite{Odlyzko2,Odlyzko3,ShapiroV}.)
Unfortunately this will increase complaints about unfairness
\cite{Odlyzko3}.  Many of the prices and policies will seem arbitrary.
That is because they will be largely arbitrary, designed to make
customers pay according to their willingness and ability to pay.  The
current U.S. airline pricing practices are a good example of the
practices that work well in providing service to a wide spectrum of
users with varying needs.  However, those practices are universally
disliked.  That may also be the fate of scholarly journal publishing
in cyberspace.

Pricing according to value means different prices for different
institutions.  Hollywood rents movies to TV networks at prices
reflecting the size and affluence of that network's audience, so that
a national network in Ireland will pay much more than that of Iceland,
but much less than one of the large U.S. networks.  We can expect
prices of electronic scholarly journals to be increasingly settled by
negotiations.  The consolidation of publishers as well as libraries
(through consortia) will help make this process manageable.

There is unhappiness among scholars and librarians about restrictions
on usage of some electronic databases, such as limiting the number of
simultaneous users, or restricting usage to a single workstation
inside the library.  The preferred method of access is, of course,
from the scholar's office.  However, that is precisely the point; to
offer a more convenient version (such as one available without
restrictions from any place on campus) for a high price, and a less
convenient version (that requires a physical visit to the library, and
possibly waiting in line) for a lower price.  Such techniques are
likely to proliferate, and a natural function for libraries will be to
enforce restrictions imposed by publishers.  We can already see this
in the license conditions for hybrid journals that appear both in
print and electronic formats.  Publishers of such journals almost
universally allow only the print version to be used for interlibrary
loans.  Although no publisher has explained clearly the rationale for
this restriction, it is easy to figure out its role.  Obtaining a copy
of the paper article is slow, cumbersome, and expensive, and this
serves to deter wide use of interlibrary loans as substitutes for
owning the journal.  If interlibrary loans of electronic versions were
allowed, though, the borrower would be in almost the same position as
a subscriber.  Even if only paper copies of electronic versions of an
article were allowed, the ease of making the copy from the digital
form and mailing it out would make interlibrary loans much faster and
less expensive, and that might undermine the market for subscriptions.

Artificial restrictions in order to maintain subscriptions are
becoming much more obvious in cyberspace than in print, but are not
new.  For example, even a casual examination shows that the Copyright
Clearance Center (CCC) and the copyright litigations of the last two
decades have practically no economic value to publishers aside from
restricting photocopying and thus maintaining the subscriber base.  In
the fiscal year ending June 30, 1997, CCC paid \$35 M to copyright
holders from the fees it collected.  Not all this money was for
scholarly publishing, and even if it were, it is tiny compared to
total revenues in the U.S. for scholarly publishers, which amount to
several billion dollars per year.  Thus all the legal attacks on
supposedly illicit photocopying and the demands for CCC fees provide
little additional revenue.  However, they do serve to discourage
dropping of subscriptions, by making copying more expensive and
more cumbersome.

Many scholars have run into problems with obtaining permission to
republish their works in collected papers volumes and the like, with
reprint fees often being demanded.  Yet such fees bring in trivial amounts of
money.  Some publishers, such as the American Economic Association
\cite{Getz} and ACM, grant blanket permissions for copying for educational
use, as they have decided that the costs of handling all the copy
requests were higher than the revenue derived from that activity.
Thus this is another case of a barrier that exists not to increase
revenues directly, but to discourage copying.

A major concern of librarians and scholars alike is that publishers
will move towards a ``pay-per-view'' model \cite{Kiernan2}.  There is little
evidence of this happening, and on balance, just the opposite is
occurring.  There is spread of consortium licensing, in which a
publisher licenses all its electronic journals to all the institutions
in a region, state, or even country (with the United Kingdom taking
the lead in national licensing).  This was to be expected.  While
there are some economic models that favor pay-per-view \cite{ChuangS}, and
such pricing approaches are likely to be used in some fraction of
cases, to deal with unusual needs, subscriptions, bundling, and site
licensing are likely to dominate.  This conclusion is supported by
standard economic models (cf. \cite{BakosB,Odlyzko3,Varian}).  It is
also supported by empirical evidence of people's aversion to
pay-per-view (cf. \cite{FishburnOS}) and by estimates of scholars'
willingness to pay for information as individuals \cite{Hunter,Odlyzko1}.

There are likely to be ``pay-per-view'' options, but they will probably
be of marginal importance, just for dealing with demand from those who
do not fit into the large classes covered by some subscription or
site-license model.  A major reason for this is ``sticker shock.''
Recall that the typical article brings its publishers revenues of
about \$4,000.  On the other hand, all studies that have been carried
out suggest that such an article is read, even if superficially (i.e.,
going beyond just glancing at the title page and abstract) by a couple
of hundred scholars.  This is also consistent with data from the
Ginsparg archive, where on average a paper is downloaded on the order
of 150 times in its first two years there.  If we assume 200 readers,
then to obtain the current \$4,000, the charge for ``pay-per-view'' would
have to be \$20.  I predict that few scholars would be willing to pay
that much, especially for an article they had only seen the abstract
for, even if the money came from their grants or departmental budgets.
Of course they effectively do pay that much now, but the charges are
hidden.  (In fact, their institutions are paying \$60 for each article
read, of which \$20 goes to the publisher, and \$40 to internal library
costs.)  A shift to ``pay-per-view'' would expose the exorbitant costs
of the current system.

Bundling, site licensing, and consortium pricing are all strategies
that enable publishers to increase their revenues by averaging out the
different valuations that separate readers or libraries place on
articles or journals.  Many librarians regard consortia as
advantageous because they supposedly provide greater bargaining power
and thus lower prices.  However, they are more likely to be helpful to
publishers in maximizing their revenues.  Consider a simple example of
a library consortium formed by three institutions, call them A, B, and
C. Suppose that A is a major research university, B a big liberal arts
school with some research programs, and C a strictly teaching school.
Consider a publisher of the (fictional) {\em Journal of Zonotopes} (JZ).
Suppose the annual institutional subscription is \$2,000, and currently
only A receives it.  Further, suppose that B and C used to subscribe,
but stopped once the price exceeded \$1,000 a year (for B) and \$200
(for C). Thus the publisher may well conclude that B and C might still
be willing to pay \$1,000 and \$200 per year for JZ, respectively.  If
the publisher were to stick to the policy of a uniform price for each
institution, it could not gain anything by lowering JZ's price, and
would risk losing A's subscription by raising it.  Suppose that
instead the publisher offers the consortium of A, B, and C a deal in
which for a total price of \$2,500 per year, A continues to receive a
print copy of JZ, and all three schools get unrestricted access to the
electronic version.  Even if the faculty and students of schools B and
C value the electronic version of JZ at half of the print value, and
those of A place no value on the digital format, the total value of
the package to the three institutions would be \$2,600 per year, and so
collectively they would be likely to spend the extra \$500.

To pursue the example above in greater detail, let us note that the
attractiveness of the consortium offer is much greater than presented
above if one also considers internal library costs.  Institution A is
really valuing JZ at \$6,000 or more, since those are its total costs
associated with the journal, while B and C value it at \$3,000 and
\$600, respectively.  Thus (even ignoring possible savings that A could
realize by dropping its print version), the consortium of A, B, and C
might be willing to pay \$3,000 or more for the package.  There are
costs associated with negotiating the license, providing assistance in
accessing the electronic version of the journal, and so on, but those
costs are far smaller than those associated with handling physical
collections.

The low marginal costs of providing digital information makes it
possible to distribute that information widely.  If some benefactor
offered to purchase for Smith College, say, all the materials that
Harvard acquires, this would bankrupt Smith, as it would not be able
to pay for proper handling of the huge mass of material.  On the other
hand, an offer of electronic access to all the materials that Harvard
has access to could be provided inexpensively.  What we are likely to
see with the spread of library consortia is much wider access to
information than we ever had before.  National licensing plans are the
extreme example of this, with everybody inside a country getting
access to all of a publisher's material.

Bundling is likely to be widespread.  Several publishers already offer
their electronic journals in a single package, with no chance for
purchasing access to a subset.  This minimizes administrative costs,
but more important, again helps take advantage of uneven preferences
for different journals to obtain higher revenues.  It also has the
advantage of protecting publishers from the subversive influence of
preprints.  Several areas, and theoretical high energy physics in
particular (since it has relied on the Ginsparg archive the longest),
might already be willing to give up most of their journals, if hard
economic times came, and academic decision makers came to departments
with offers of the type ``Either you give up your journals, or you give
up three postdocs.''  In most areas, though, such a move is not
feasible, since the preprint culture is not sufficiently developed.
Now if the journals in theoretical high energy physics only come in a
package with other journals from less advanced fields, then an offer
like that above cannot be made.  Thus bundling can serve the
publishers' economic interests in retarding evolution of scholarly
publishing to the rate of the slowest area.

Scholarly publishers are consolidating, with Elsevier, already the
largest player in this market, in the forefront of the acquisition and
merger wave.  The publishers' market power may be counterbalanced,
though, by the rise of library consortia.  How the publisher oligopoly
will interact with purchaser cartels will be an interesting phenomenon
to watch.

\section{Will it work?}
\hsp
Will the publishers succeed in disintermediating the libraries, and
preserving their revenues?  There are two problems they face.  One is
a short-term one.  While electronic publication will eventually reduce
the expenses of both publishers and libraries, right now it is raising
those expenses, as both parties have to handle print and digital media
at the same time.  The other problem, the longer-term one, is that
publisher revenues are far greater than is necessary to provide
quality sufficient for primary publications.  The manuscripts prepared
by authors have been improving, to the point that all the copy editing
and typesetting that publishers contribute is of diminishing value.
Furthermore, in spite of the attempts of some publishers, there is no
way to stop the preprint tide.  The free circulation of preprints
offers so many advantages to scholars that it is only a matter of time
until they become universal.  To survive in the long run, publishers
will have to contribute more that is of real value.  They are starting
to do so by adding links to their electronic articles and similar
measures.  I suspect they will have to do a lot more.  Until they do,
they are vulnerable.  Their main danger will come not from competition
by Kinko's, but from a change in perceptions by administrators.

The analogy with {\em Encyclopaedia Britannica} might serve to illuminate
the danger.  To quote from \cite{EvansW} again,
\begin{quote}
  Judging from their initial inaction, Britannica's executives
  failed to understand what their customers were really buying.
  Parents had been buying Britannica less for its intellectual
  content than out of a desire to do the right thing for their
  children.  Today when parents want to ``do the right thing,''
  they buy their kids a computer.
\end{quote}
Nontraditional methods for information dissemination (preprints, but
also email, Web pages, and so on) are growing in importance.  At some
point the administrators in charge of libraries may decide that ``doing
the right thing'' for their faculty and students means redirecting
resources away from traditional expensive journals.

\paragraph{Acknowledgements:}
I thank Stevan Harnad and the the other members of
the American Academy of Arts and Sciences study group on transition
from paper for their comments.

\end{document}